\documentstyle[epsf]{article}

\begin{document}

\title{Numerical simulation of a binary communication channel:
Comparison between a replica calculation and an exact solution }

\author{
D.R.C. Dominguez\thanks{Present address: Instituut voor 
Theoretische Fysica, K.U. Leuven, 3001, Leuven, Belgium.},
M. Maravall\thanks{Present address: Institute for Theoretical 
Physics, State University of New York at Stony Brook, 
NY 11794-3840, USA.},
A. Turiel,
J.C. Ciria\thanks{Permanent address: Departamento de 
F\'{\i}sica Te\'orica. Universidad de Zaragoza.C/ Pedro Cerbuna, 12   
50010 Zaragoza. SPAIN} 
and N. Parga
\\
\\
Departamento de F\'{\i}sica Te\'orica,
Universidad Aut\'onoma de Madrid,  \\
Cantoblanco, 28049 Madrid, Spain
}

\maketitle

\begin{abstract}
The mutual information of a single-layer perceptron with $N$ Gaussian
inputs and $P$ deterministic binary outputs is studied by numerical
simulations.  The relevant parameters of the problem are the ratio
between the number of output and input units, $\alpha = P/N$, and
those describing the two-point correlations between inputs.  The main
motivation of this work refers to the comparison between the replica
computation of the mutual information and an analytical solution valid
up to $\alpha \sim O(1)$.  The most relevant results are: (1) the
simulation supports the validity of the analytical prediction, and (2)
it also verifies a previously proposed conjecture that the replica
solution interpolates well between large and small values of $\alpha$.

\end{abstract}


Pacs:

87.10.+e: General, theoretical, and mathematical biophysics

64.60.Cn: Order-disorder transformations; statistical mechanics of 
model systems

{\it Europhysics Letters} {\bf 45}{6}, 739-744 (1999)

\section{Introduction}

The extraction of sensory information by the brain from a stream of
multi-dimensional data may be understood as a process of optimisation
of mutual information $(MI)$ \cite{linsker88} or of redundancy
\cite{Ba61}.  The $MI$ measures the statistical dependence between two
random variables \cite{SW49}.  In our case, they correspond to an
$N$-dimensional input signal $\vec{\xi}$ provided by the sensory
receptors, and a $P$-dimensional output $\vec{v}$.  More precisely,
the $MI$ indicates the amount of knowledge about $\vec{\xi}$ that can be
extracted from $\vec{v}$ (See e. g. ref.\cite{SW49}).  With respective
probability densities $p_{\xi}$ and $p_{v}$, this is given by

\begin{equation}
I[p_{v},p_{\xi}] =
<\log {p_{v,\xi}(\vec{v},\vec{\xi})\over 
p_{v}(\vec{v})p_{\xi}(\vec{\xi})}>_{v,\xi},
\label{1.Ip}
\end{equation}
here
$\log(x) \equiv \frac{\ln (x)}{\ln 2}$. 
The average $<...>_{v,\xi}$ is over the joint
probability distribution $p_{v,\xi}(\vec{v},\vec{\xi})$.
For instance, if $\vec{v}$ and $\vec{\xi}$ are independent,
we have $p_{v,\xi} = p_{v}\cdot p_{\xi}$
and so $I=0$.

The problem of learning the statistical properties of a set of
$N$-dimensional correlated Gaussian inputs is well understood for a
linear channel, even in the presence of noise (which is actually
necessary to regularise $MI$)\cite{Li92}-\cite{CG95}.  A non-linear
continuous channel has also been studied in the low-noise limit for
rather general transfer functions \cite{NP94}.  It was shown that
maximisation of the $MI$ leads to a factorial code.  Threshold-linear
networks \cite{Treves95,Schultz98} (treated with the replica
technique) have also been considered.  On the other hand, the binary
channel, where the outputs take discrete values (say $v_{\mu}=\pm 1$),
is not well understood, although the problem has been studied using
replica-symmetric ($RS$) statistical-mechanical techniques
\cite{NP93}-\cite{TKP98}.  Most interestingly, an analytical solution
has been found, and the existence of a large order phase transition as
the number of output neurons increases has been suggested
\cite{TKP98}.  The relevant parameter to describe this transition's
occurrence is the ratio between the number of output and of input
units, $\alpha = P/N$.  The analytical solution holds up to a value of
$\alpha$ of order one, beyond which it is not longer correct. On the
contrary, the $RS$ solution does not exhibit any transition and gives
good approximations at both the small and large $\alpha$ regimes. It has
been proved that below some $\alpha \sim 1$ the analytical and the
$RS$ solutions are very close. In fact, an expansion in powers of
$\alpha$ shows that the two solutions are identical up to
$O(\alpha^2)$. In spite of the fact that from the third order the 
corresponding expansions differ, the numerical agreement up to 
$\alpha\sim 1$ is excellent (a relative difference of less than 0.9\% up 
to $\alpha=0.1$). This is due to intriguing cancellations between higher 
orders.

Here we present simulations with the aim of providing numerical
evidence on the validity of the analytical solution.  In addition, since
the order of the transition is large and the $RS$ solution seems to
interpolate well between the small $\alpha$ and the asymptotic
regimes, we also compare numerical simulations at several values of
$\alpha$ with the replica theory prediction \cite{NP93}.

\section{The Binary Channel}

We consider a single-layer perceptron, or $channel$,
with $N$ continuous input neurons whose states
$\xi_{i}$ define a vector
$\vec{\xi}=\{\xi_{i}\}_{i=1}^{N}$
representing the $signal$ received from the environment.
The output layer has $P$ binary neurons 
the values $v_{\mu}=\pm 1$ of which compose the vector
$\vec{v}=\{v_{\mu}\}_{\mu=1}^{P}$,
that represents the $code$. 
Between the signal and the code there is an encoder,
given by a set of synaptic couplings
${\cal J} \equiv \{ {\cal J}_{i\mu} \}$. 

The inputs take values drawn from an
$N$-dimensional Gaussian probability distribution,
unbiased ($<\xi_{j}>=0$) and with correlation matrix 
${\bf C}_{X}^{ij} = <\xi_{i}\xi_{j}>$,

\begin{equation}
p_{\xi}(\vec{\xi}) =
e^{-\vec{\xi}\cdot {\bf C}_{X}^{-1} \cdot\vec{\xi}/2}
/\sqrt{ 2\pi C_{X} },\, 
C_{X}=|{\bf C}_{X}|,
\label{2.px}
\end{equation}
which using a convenient shorthand can be expressed as $\vec{\xi}
\doteq N(\vec{0},{\bf C}_{X})$.  (Here $X$ is a correlation parameter
between input neurons, to be defined more precisely later).
The transfer function is deterministic, so that $v_{\mu} = {\mbox{sign}} 
(h_{\mu})$, where

\begin{equation}
h_{\mu} = \sum_{i=1}^{N} J_{i\mu} \xi_{i}
\equiv \vec{J}_{\mu}\cdot\vec{\xi}
\label{2.vm}
\end{equation} 
and the $\vec{J}_{\mu}$'s denote the
synaptic weight vectors linking the signal $\vec{\xi}$ to each output 
neuron $\mu$. They form a set of independent random vectors
$\{\vec{J}_{\mu}\}_{\mu=1}^{P}$, 
each distributed according 
to an $N$-dimensional Gaussian probability, 
with mean $<J_{i\mu}>=0$ and correlation matrix 
${\bf \Gamma_{\mu}}$, whose elements are
$\Gamma_{\mu}^{ij} = <J_{i\mu}J_{j\mu}>$. This means that 
$\vec{J}_{\mu} \doteq N(\vec{0},{\bf \Gamma_{\mu}})$.

We are mainly interested in computing the averaged mutual information
per input unit in the thermodynamic limit, {\it i.e.} 
$i\equiv \lim_{N \rightarrow \infty}{1\over N} <I[p_{v},p_{\xi}]>_{\cal{J}}$.  A
useful result to have in mind is that the $MI$ in eq.(\ref{1.Ip}) is
the difference

\begin{equation}
I[p_{v},p_{\xi}]_{\cal{J}} =
H[p_{v}] - H[p_{v|\xi}],
\label{2.Ip}
\end{equation}
where $H[p_{v}]= -<\log p_{v}>_{v}$ is the entropy of the output, while
$H[p_{v|\xi}]= -<<\log p_{v|\xi}>_{v|\xi}>_{\xi}$ is the conditional
entropy of the output given the input (averaged over the input).  The
code $\vec{v}$, given a fixed signal $\vec{\xi}$, has the conditional
probability $p_{v|\xi}= p_{v,\xi}/p_{\xi}$.  
Since a deterministic channel
clearly has zero conditional entropy, in that case the $MI$
reduces to the output entropy.

\section{Analytical results for an example}

As an example we consider a Gaussian input distribution with 
two-point correlation ${\bf C}_{X}$ given by $C_{X}^{ij}=X^{|i-j|}$.
The input neurons are then less (more) spatially correlated if $X\ll
1$ ($X\sim 1$).  For $X=1$, ${\bf C}_{X}$ is a singular matrix, while
for $X\to 0$ ${\bf C}_{X}$ tends towards the identity matrix.  The
correlations between two synapses converging to the same output,
$\Gamma^{ij}_{\mu}$, are chosen to be equal for all output neurons and
normalised to $1$, $i.e.$ $\Gamma^{ij}_{\mu}=\Gamma^{ij}= 
\delta_{ij},\,\forall \mu$.

For small $\alpha$, the MI can be expanded as  
$i(\alpha)\sim a_{0}+a_{1}\alpha+a_{2}\alpha^{2}+a_{3}\alpha^{3}$.
One trivially obtains that $a_{0}=0$ and $a_{1}=1$. Both the $RS$ 
solution (see \cite{NP93}) and the analytical solution (see \cite{TKP98}) 
give the same value of $a_{2}$,

\begin{equation}
a_{2}(X)= {-\gamma\over\pi^{2}\ln(2)} \\
\label{4.a2}
\end{equation}

\noindent
where $\gamma\equiv {1+X^{2}\over 1-X^{2}}$.
On the other hand, the two techniques disagree in their predictions for $a_{3}$. 
The $RS$ method yields

\begin{equation}
a_{3}^{RS}(X)= {(6-12/\pi)\gamma^{2}-2\over 3\pi^{3}\ln(2)},
\label{4.a3}
\end{equation}
while (following the methods of \cite{TKP98}) one can easily verify
that the  analytical technique gives:

\begin{equation}
a_{3}^{An}(X)= {6\gamma^{2}-2\over 3\pi^{3}\ln(2)}.
\label{4.aA}
\end{equation} 

The $RS$ solution also gives rise to predictions for strongly correlated 
inputs. For $X\sim 1$ one obtains:

\begin{equation}
i\sim K(1-X)^{1/3}\alpha^{\nu},\,\nu\equiv 2/3.
\label{4.iK}
\end{equation} 

\noindent
Finally, the same solution shows logarithmic behaviour for large $\alpha$:

\begin{equation}
i\sim\log\alpha + {1\over 2\ln 2}+ \log 0.72 + {1\over 2N} {\mbox{Tr}}[\log
{{\bf C_{X}}\over N}].
\label{4.il}
\end{equation}

\begin{figure}[htb]
\begin{center}
\hspace*{0cm}
\epsfxsize=5.cm
\epsfbox[100 100 470 380]{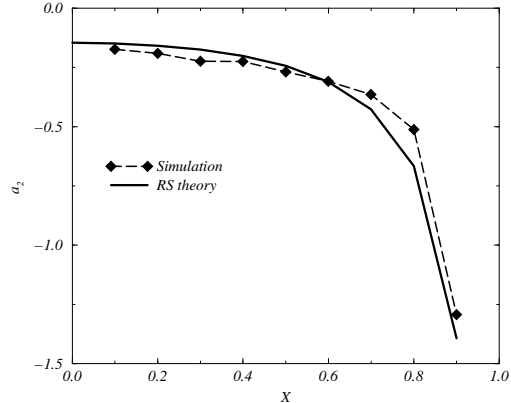}
\end{center}
\caption{The coefficient $a_{2}(X)$ obtained from simulations 
compared with the $RS$ prediction.}
\label{1ax}
\end{figure}

\section{Method}


The first step of the simulation entails choosing a coupling sample
$\cal{J}$ at random.  A given choice of $\cal{J}$ will be labelled as the 
sample $s$, and the total number of samples will be denoted by $S$.  
Next, signals
$\vec{\xi}$ are drawn in order to obtain several codewords $\vec{v}$.
A histogram $p^{s}_{v}(\vec{v})$ is then constructed from these
states, representing the probability $p_{v}(\vec{v})$ and
allowing us to estimate the output entropy:

\begin{equation}
i_{s} = 
- {1 \over N} \sum_{\{v_{\mu}=\pm 1\}} p^{s}_{v}\log p^{s}_{v}.
\label{3.Hs}
\end{equation}
Clearly, this approximation to the true code entropy $H[p_{v}]$ will
improve as the number of drawn signals increases.
In practice, we evaluated $p^{s}_{v}(\vec{v})$ using about $100$
different input states $\vec{\xi}$.
The final step is to calculate the average over an ensemble of
$\cal{J}$, to get $i= {1 \over S} \sum_{s=1}^{S}  i_{s}$.  


\section{Results from simulations}
\label{IV}

Here we present studies of
three relevant regions in parameter space $(\alpha,X)$:
(1) $\alpha$ small ;
(2) intermediate values of $\alpha$ and strong
correlations, $X\sim 1$; and (3) $\alpha$ large.
The results are compared with the theoretical predictions,  
eqs.(\ref{4.a2})-(\ref{4.il}).  The self-averaging property of the MI is 
also analysed.


In the {\bf small $\alpha$ case}, we calculated $i(\alpha)$ for values of 
$\alpha$ ranging from $0.05$ to $0.2$: we fixed $P=10$ and took a 
variable number of inputs in the interval $N=50,...,200$. We repeated this 
process for several values of the correlation parameter $X=0.1,0.2,...,0.9$. 
Using that $i(\alpha=0)=0$ when $N\to\infty$, we performed a linear 
regression on $i$, $i(\alpha,X)/\alpha=a_{1}(X)+a_{2}(X)\alpha$, and obtained the 
coefficients $a_{1}, a_{2}$ as a function of $X$.
For all values of $X$, $a_{1}\sim 1$, which is in agreement with the theory.
The function $a_{2}(X)$ is plotted in fig.\ref{1ax}, in comparison with 
the theoretical prediction (eq.(\ref{4.a2})). The good agreement indicates that the 
thermodynamic limit solution is a good estimation even for $N$ not very 
large, as long as one is in the low-loading expansion.

\begin{figure}[htb]
\begin{center}
\hbox{
	\makebox[12.25cm]{$a_2$\hspace*{2.5cm}$a$\hspace*{3.5cm}$a_3$\hspace*{2.5cm}$b$\hspace*{2.5cm}}
}
\vspace*{0.5cm}
\hbox{
	\makebox[12.25cm]{
		\hspace*{0.5cm}
		\leavevmode
		\epsfxsize=5.5cm
		\epsfysize=2.75cm
		\epsfbox[70 100 400 280]{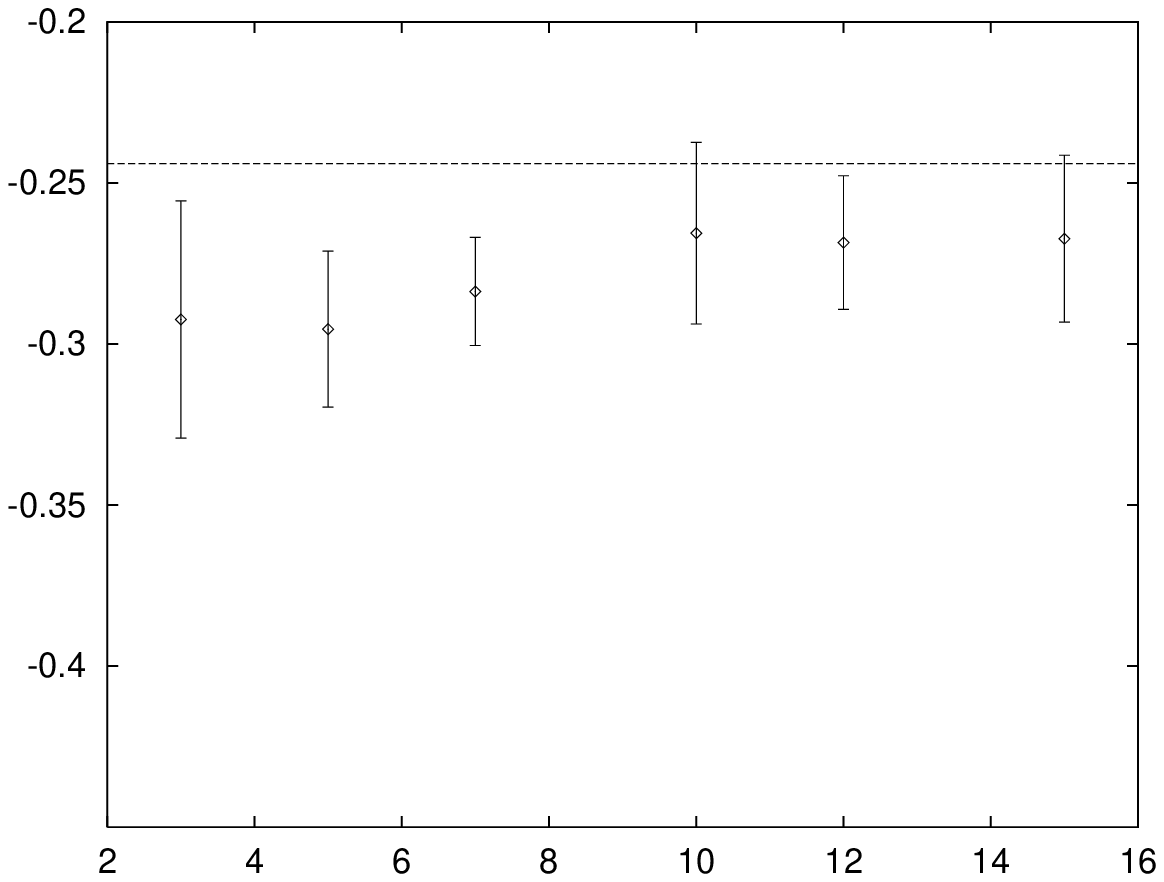}
		\hspace*{0.75cm}
		\epsfxsize=5.5cm
		\epsfysize=2.75cm
		\epsfbox[70 100 400 280]{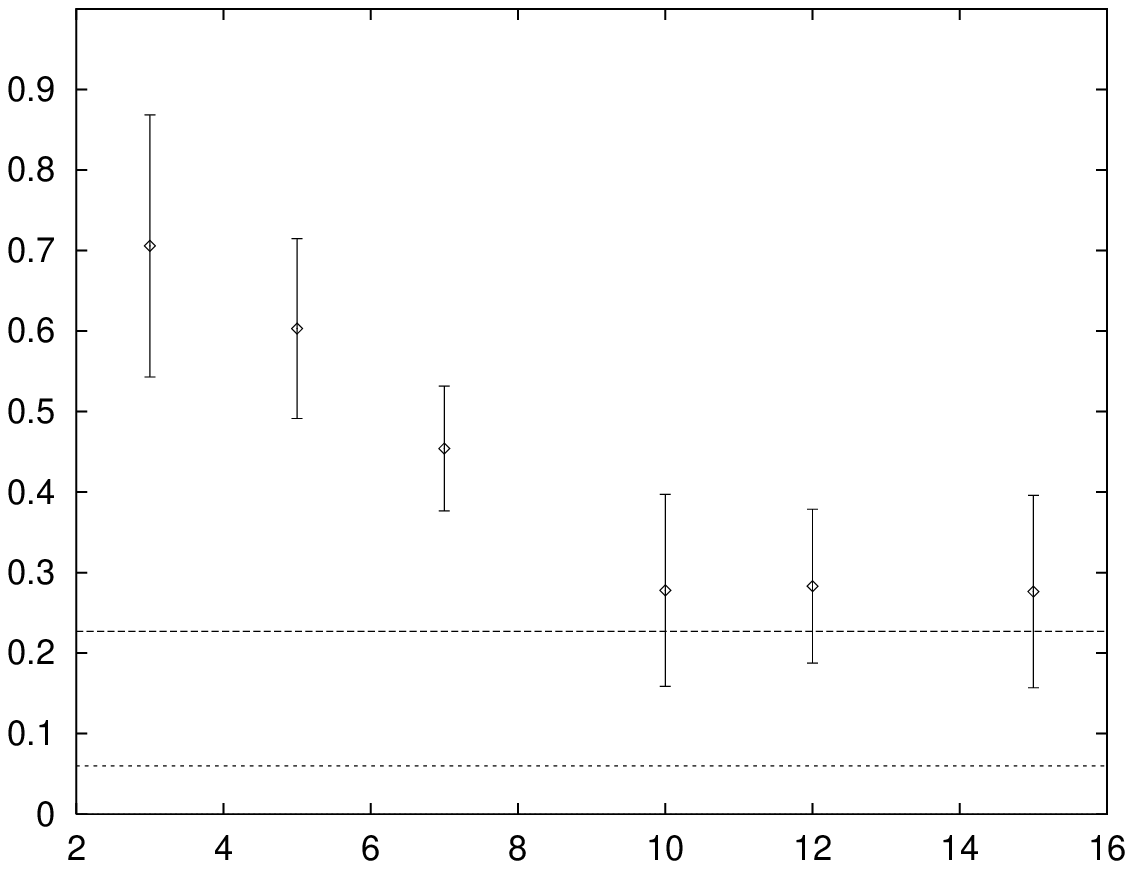}
	}
}
\end{center}
\hbox{
	\makebox[12.25cm]{\hspace*{1.25cm}$P$\hspace*{6cm}$P$}
}
\caption{ The coefficients $a_2$ (a) and $a_3$ (b) for
several $P$'s, for $X=0.5$. The straight horizontal lines 
represent the theoretical values; $a_2^{RSA}=a_2^{An}=-0.244$ and 
$a_3^{RSA}=0.06$, $a_3^{An}=0.23$. The exact solution is in good 
agreement with the data, while the RSA coefficient lies outside the error 
bars. } 
\label{2ax}
\end{figure}

\indent
In order to evaluate $a_3$, we set $a_{1}\equiv 1$ (to 
avoid increasing the error through a larger number of parameters to be fitted).
We analysed the MI only for the value $X=0.5$. From eq. \ref{4.a2} 
the theoretical value $a_2=-0.244$ can be obtained, and from eqs. \ref{4.aA} and 
\ref{4.a3} we have $a^{An}_{3}=0.227$ and $a^{RS}_{3}=0.063$ respectively. 
The simulation itself was done using several values of $P$, and for each  
of them a linear regression was performed using
$[i(\alpha)/\alpha-1]/\alpha=a_{2}+a_{3}\alpha$. 

\indent
In this case the corrections due to finite 
size effects are noticeable. To observe convergence to the 
asymptotic regime we considered several values of $P$, namely 
$P=3,5,7,10,12$ and $15$. We averaged for each over, respectively,  
$S=2500,1000,200,100,50$ and $20$ samples of $\cal{J}$. The number of 
samples $S$ was taken larger as the number of outputs $P$ became smaller, to 
compensate for the lack of statistics. 
Numerical evaluation of $a_3$ is extremely costly in 
computational time; this is the reason why we restricted ourselves 
to a single value of $X$ and did not consider values of $P$ 
larger than $15$.

\indent
The result is that 
our simulation is in agreement with the analytical prediction. 
As we see in fig.\ref{2ax}b, $a_{3}^{P}$ converges to $a_{3}\sim 0.28$ as $P$ 
increases. Given the error bars, this result is compatible with 
$a^{An}_{3}$ but {\em excludes} $a^{RS}_{3}$. For the sake of 
comparison we have included fig. \ref{2ax}a where the same numerical analysis is done 
for   $a_{2}$.

\begin{figure}[bht]
\begin{center}
\hbox{
	\makebox[15cm]{a)\hspace*{5cm}b)\hspace*{5cm}c)}
}
\vspace*{0.75cm}
\hbox{
	\makebox[15cm]{
		\leavevmode
		\epsfxsize=5cm
		\epsfysize=3.5cm
		\epsfbox[35 70 553 385]{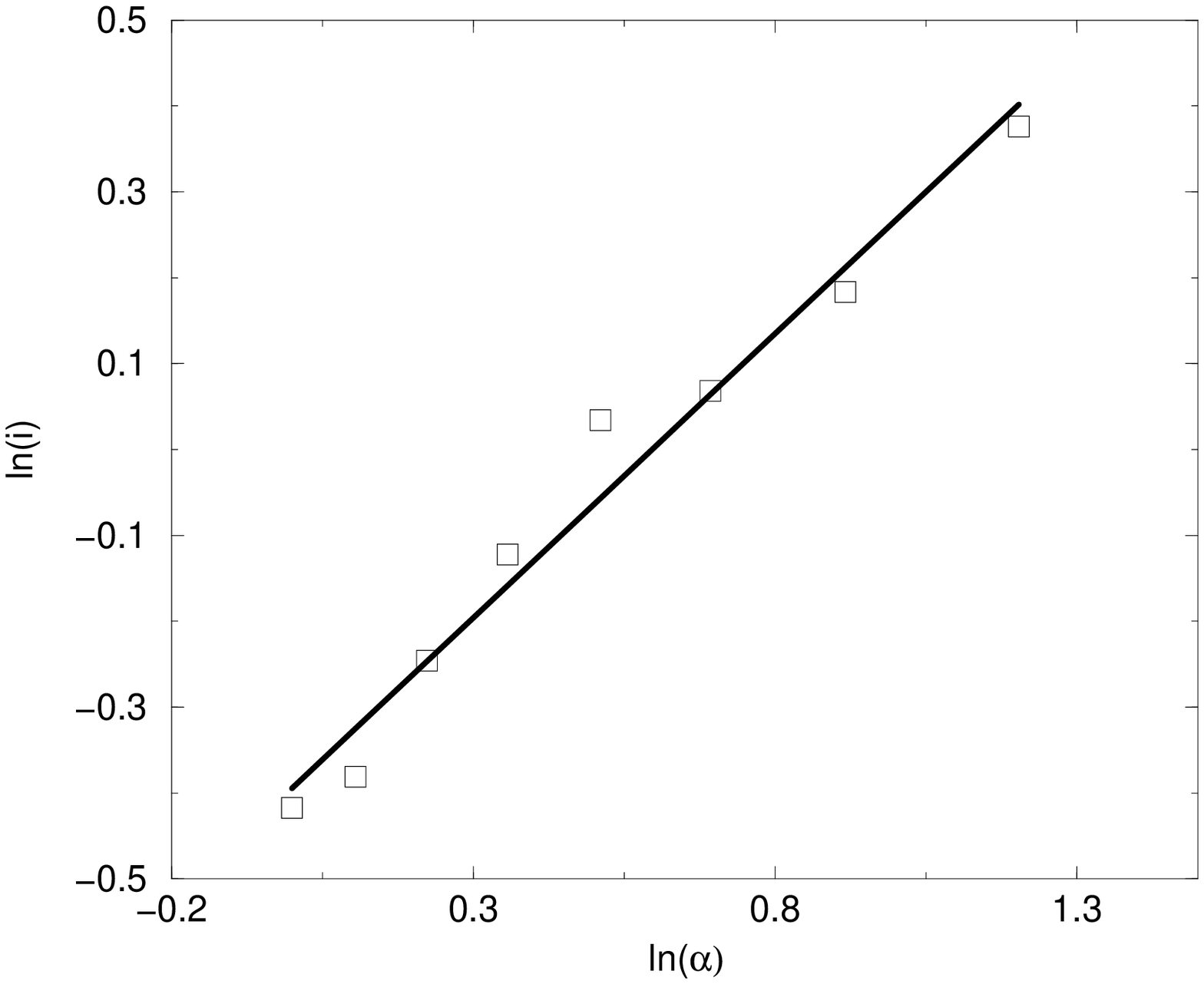}
		\epsfxsize=5cm
		\epsfysize=3.5cm
		\epsfbox[35 70 553 385]{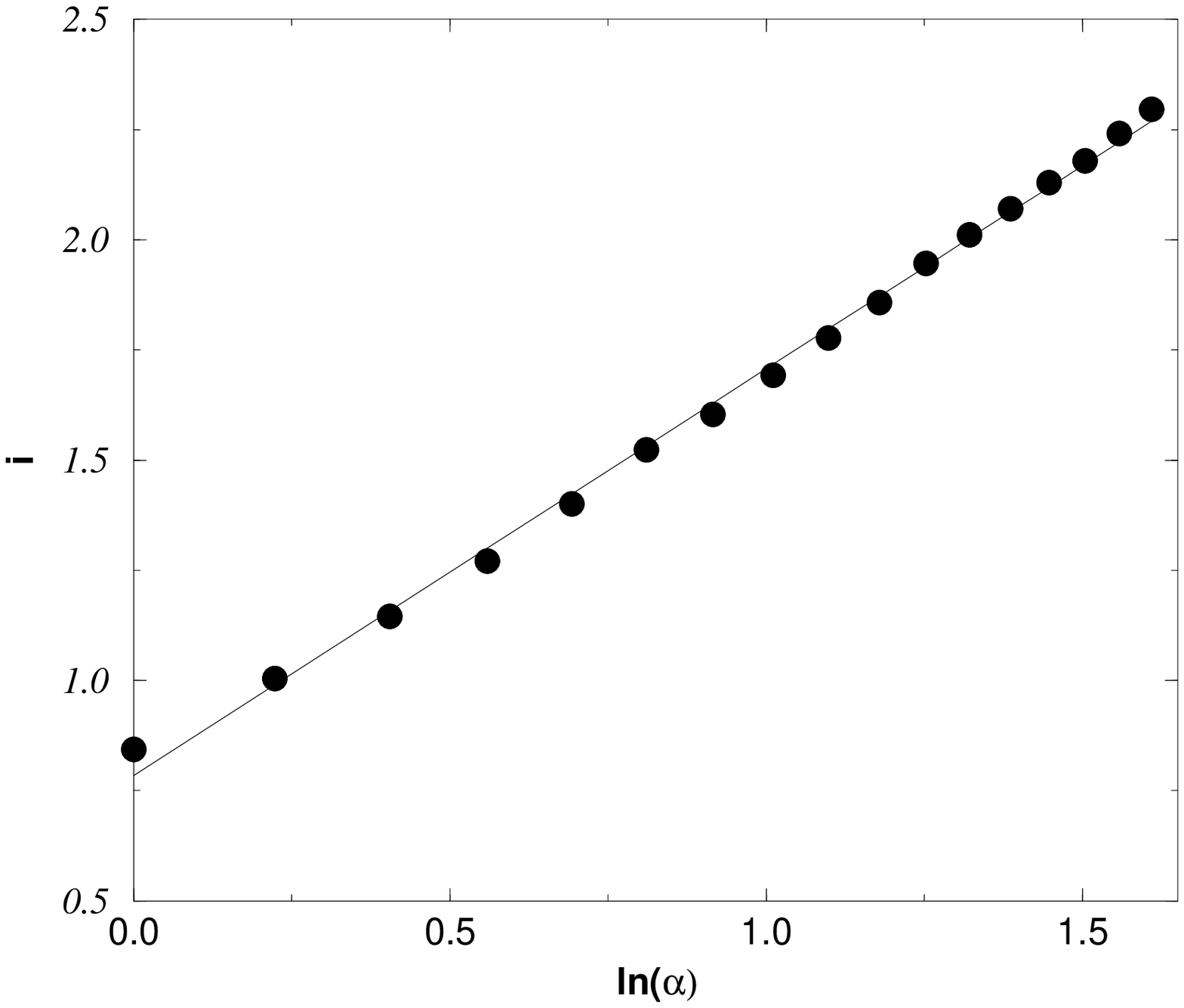}
		\epsfxsize=5cm
		\epsfysize=3.5cm
		\epsfbox[35 75 553 420]{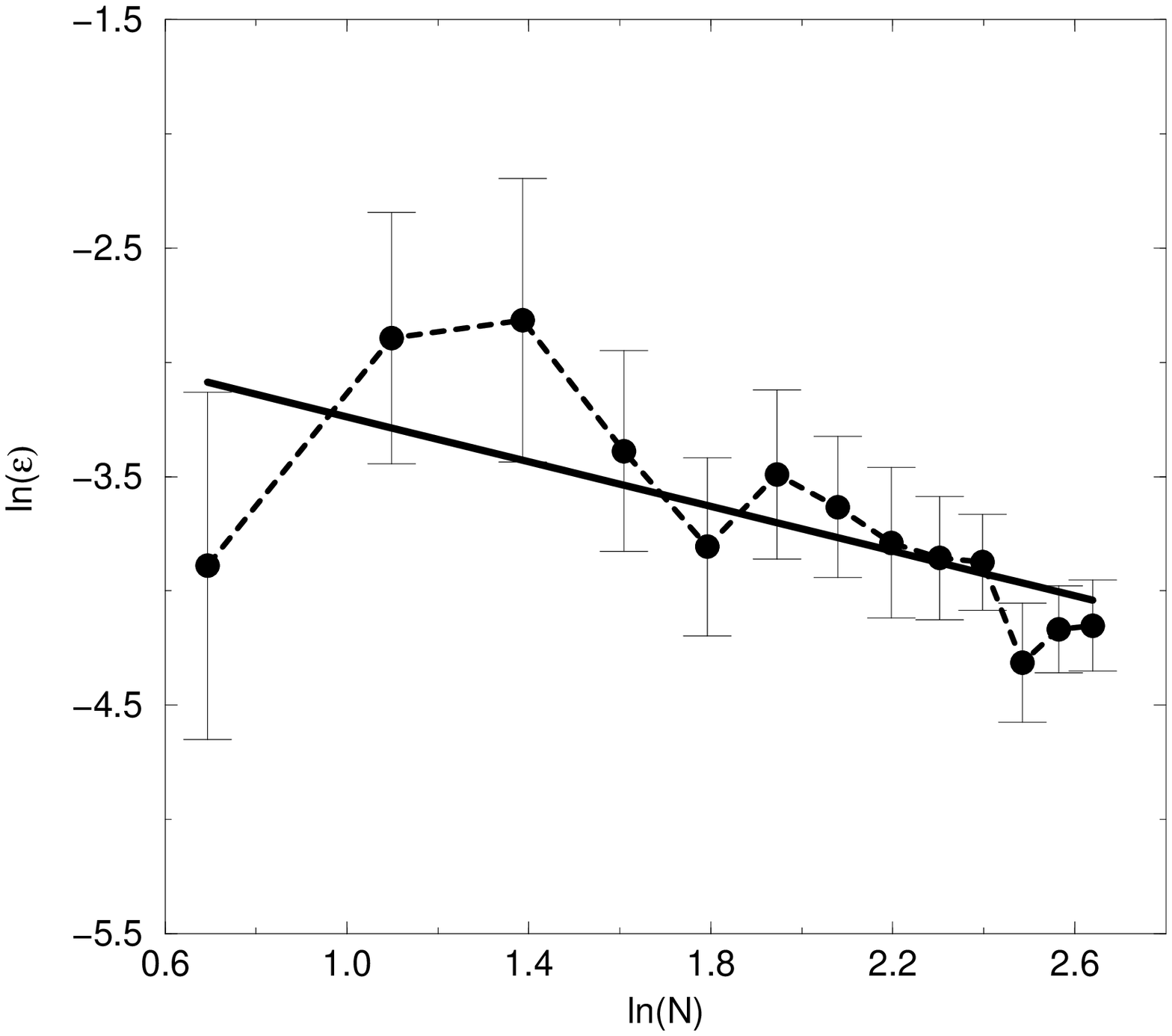}
	}
}
\end{center}
\vspace*{-0.3cm}
\caption{ {\bf a)} The behaviour of the $MI$ for 
strongly correlated inputs ($X=0.9$). 
$P=10$ and $N=3,...,10$. 
The squares are obtained from simulation,
while the plotted line is: 
$\ln(i)=-0.395+0.662\ln(\alpha)$. 
{\bf b)} The $MI$ for large $\alpha$,
with $N=4, P=1,...,20$ and $X=0.1$ ($\alpha\leq 5$).
The dots are obtained from simulation, 
while the plotted curve  is
$i=0.78+0.92\ln(\alpha)$. 
{\bf c)} The self-averaging property.
We took $X=0.9$ and $P = N = 2,...,14$ ($\alpha = 1$).
The linear regression gives
$\ln[\Delta(i)/i]= -2.5-0.49\ln(N)$, where
$\Delta^{2}(i)= 
{1\over S} \sum_{s=1}^{S} [i_{s}-i]^{2}$. 
}
\label{3ix,il,is}
\end{figure}


\indent
The results for {\bf strongly correlated inputs} are shown 
in fig. \ref{3ix,il,is}a. There is good agreement between the 
simulation and the exponential behaviour of eq.(\ref{4.iK}): $\nu= 
0.662\sim 2/3$.


The results for the {\bf large $\alpha$ limit} are presented in
fig.\ref{3ix,il,is}b.  The logarithmic behaviour predicted in
eq.(\ref{4.il}) is observed.


To verify that the MI is self-averaging we calculated its mean-square
deviation $\Delta(i)$ over the samples and made a fit to the form
${1\over \sqrt{N}}$.  The good agreement with this expression can be
seen in fig. \ref{3ix,il,is}c.
This shows that the methods of statistical mechanics
are appropriate to studies of the information in binary
channels.

\section{Conclusions}

Our main result refers to the comparison between the $RS$ 
\cite{NP93} and analytical \cite{TKP98} solutions. 
The difference between them can be seen in a small-$\alpha$
expansion. Numerical simulation confirms that the two 
solutions coincide up to second order. At the next order the 
solutions are different (see eqs. \ref{4.a3}-\ref{4.aA}), 
and the simulation excludes the 
replica calculation while it is in agreement with the analytical 
one (see fig. \ref{2ax}).

We have also verified the conjecture that the $RS$ solution 
is a good interpolation between the small and the large 
$\alpha$ behaviors \cite{TKP98}.
In particular, the simulation shows that for intermediate 
values of $\alpha$ and strongly correlated inputs 
the MI behaves as $i\sim\alpha^{2/3}$ (see fig.\ref{3ix,il,is}a).  
Moreover, in fig.\ref{3ix,il,is}b we see
that the expected logarithmic behaviour for the large $\alpha$ case
fits the MI very well.

\section{Acknowledgments}
This work was supported by a Spanish grant PB 96-47.
Antonio Turiel is financially supported by an FPI grant from the Comunidad
Aut\' onoma de Madrid, Spain.
M. Maravall was supported by a Beca de Colaboraci\'{o}n from the Spanish
Ministry of Education. 
D.R.C. Dominguez thanks the K.U. Leuven
for a research fund (grant OT/94/9).

\clearpage

\end{document}